# Multimodal Functional and Structural Brain Connectivity Analysis in Autism: A Preliminary Integrated Approach with EEG, fMRI and DTI

Bogdan Alexandru Cociu, Saptarshi Das, Lucia Billeci, Wasifa Jamal, Koushik Maharatna, *Member, IEEE*, Sara Calderoni, Antonio Narzisi, and Filippo Muratori

*Abstract*—This paper proposes a novel approach of integrating different neuroimaging techniques to characterize an autistic brain. Different techniques like EEG, fMRI and DTI have traditionally been used to find biomarkers for autism, but there have been very few attempts for a combined or multimodal approach of EEG, fMRI and DTI to understand the neurobiological basis of autism spectrum disorder (ASD). Here, we explore how the structural brain network correlate with the functional brain network, such that the information encompassed by these two could be uncovered only by using the latter. In this paper, source localization from EEG using independent component analysis (ICA) and dipole fitting has been applied first, followed by selecting those dipoles that are closest to the active regions identified with fMRI. This allows translating the high temporal resolution of EEG to estimate time varying connectivity at the spatial source level. Our analysis shows that the estimated functional connectivity between two active regions can be correlated with the physical properties of the structure obtained from DTI analysis. This constitutes a first step towards opening the possibility of using pervasive EEG to monitor the long-term impact of ASD treatment without the need for frequent expensive fMRI or DTI investigations.

## I. Introduction

MODELLING the information processing mechanisms of the brain as a vast network of specialized units that interact with each other has been proven to be successful in explaining and diagnosing several neurological disorders [1]. Autism spectrum disorders are one such example that can be explained as disruptions in the normal operation of this network, often described by hypo-connectivity of long range connections and hyper-connectivity of short-range connections [2], [3], [4]. The core symptoms of an ASD consist of impairment in socio-communicative abilities and the presence of repetitive behaviour of restricted interests [5].

Owing to the fact that ASDs are not a single condition, but a broad spectrum of disorders, the treatment needs to be personalized to each individual child. Ideally, one should rely on some objective measures reflecting physiological changes in the brain in a quantitative way, for ascertaining the effectiveness of the therapy. Currently, the effectiveness of a therapy is only measured based on behavioural analysis (i.e. observing whether any behavioural improvement occurs due to the therapy), which is plagued with subjective judgement and therefore may introduce systematic bias. Some pioneering studies can be found in literature, measuring the effect of intensive behavioural treatment on the brain structure and function of ASD patients through Diffusion Tensor Imaging (DTI) [6], functional Magnetic Resonance Imaging (fMRI) [7], resting-state fMRI [8], or electroencephalography (EEG) [9]. This study is a part of the EU FP7-funded MICHELANGELO project, in which one of the objectives is to understand how brain connectivity analysis could feasibly be used for generating a set of such objective measures for characterizing the actual pathophysiological status of an autistic child [10], [11], [12], [13].

Brain connectivity is usually studied in terms of the functional and structural networks. Functional connectivity focuses on the information exchange between the different regions, describing interdependencies or patterns of correlations between their activations in time. It is fast and task-dependent, changing in the order of milliseconds [14]. On the other hand, structural connectivity refers to the underlying anatomical links (white matter tracts) that support functional connectivity [15]. This can change in time due to neural plasticity, but at a much slower pace, and it is also specific to each individual. Both can be helpful in explaining the structural and functional brain correlates of ASD. Even though the information they capture is different in nature from a biological and theoretical perspective, these two influence each other [16], [17]. Several modern brain imaging methods can be used to estimate the connectivity of the brain. The white matter tracts (structural connectivity) can be assessed

Manuscript received XX-XXX; revised XX-XXXX; accepted XX-XXXX.

The work presented in this paper was supported by EU FP7 funded MICHELANGELO project, Grant Agreement #288241. URL: www.michelangelo-project.eu/. SC was partly supported by the Italian Ministry of Health and by Tuscany Region with the grant (GR-2010-2317873), and by Bando FAS Salute Sviluppo Toscana (ARIANNA Project).

B.A. Cociu, S. Das, W. Jamal and K. Maharatna are with the School of Electronics and Computer Science, University of Southampton, Southampton SO17 1BJ, United Kingdom (e-mail: bogdancociu@gmail.com, {sd2a11, wj4g08, km3}@ecs.soton.ac.uk).

L. Billeci is with the Institute of Clinical Physiology, National Research Council (CNR), Pisa, Italy (email: lucia.billeci@ifc.cnr.it)

S. Calderoni, A. Narzisi, and F. Muratori are with the IRCCS Stella Maris Foundation, Calambrone, Pisa, Italy (email: {sara.calderoni, anarzisi, filippo.muratori}@fsm.unipi.it).

with DTI, whereas functional connectivity can be captured by fMRI, providing good spatial resolution but poor temporal resolution, or by EEG, which gives a good temporal resolution but a poor spatial resolution.

In selecting an imaging method, one has to bear in mind that the fast temporal evolution of how information from different brain regions is integrated, plays a crucial role in understanding the underlying mechanisms for disorders such as ASD [2], [3]. For this reason, the information on functional connectivity provided by fMRI is insufficient for a full and accurate characterization, and incorporating information provided by EEG is required. Another aspect warranting the use of EEG for connectivity estimation is its portability and minimum intrusion compared to fMRI recording, which cannot be carried out while the patient is engaged in a task in a natural or home environment. Recently, true pervasive EEG systems emerged, featuring dry electrodes and wireless data transmission, such as the ENOBIO system [18]. This allows the child with the disorder to have their EEG conveniently recorded at home as often as necessary. However, literature on EEG-based connectivity in autism does not show convergent findings [19], and a multimodal study is therefore needed to find a methodology that might reveal network disruptions similar to those uncovered by fMRI or DTI. Structural connectivity also plays a major role in understanding ASD [3], [20], [21]. It provides an insight into the anatomical connections of the brain of each individual patient, can also aid in tailoring the treatment to different individuals and in assessing to what degree significant neural changes have occurred after long-term treatment. Ideally, EEG-derived functional connectivity could be used to capture the same information, the structural connectivity can provide. Other multimodal studies of ASDs can be found in literature [22], [23], [24], but their goal is to draw a complete picture of the disease, making use of different imaging methods, rather than extracting the most relevant information using only EEG which this paper aims at. Correlations between structure and function based on fMRI and DTI have also been reported in [25], [26], [27], encouraging such a pursuit. To the best of our knowledge there is no study, which has applied a multimodal approach based on the combination of EEG, fMRI and DTI in ASD children.

The aim of the multimodal study presented in this paper is to develop a suitable methodology for estimating functional connectivity in ASD children from pervasive EEG, in such a way that it also tracks the underlying structural connectivity. To this end, we evaluate different methods for estimating functional connectivity by the degree to which they correlate with the structural connectivity. If such a model for mapping the functional connectivity onto the structural connectivity features could be successfully made, it would help in obtaining an estimate of the underlying structural characteristics of the brain network e.g. the number of DTI tracts or tract volume etc. Once the coordinates of the relevant nodes in the network would have been supplied from fMRI information, only non-invasive and low-cost EEG measurements would be required to track the evolution of this network, followed by offline data processing. This shows the potential to curtail the application of expensive neuroimaging techniques multiple times on a child's brain to investigate incremental improvements during and after a therapy.

The experiment to develop the methodology requires fMRI, DTI and EEG to be recorded for each subject and during resting state in order to activate a well-known network structure in the brain called the default mode network. The first step in the methodology is to perform EEG source localization, to recover the source activities inside the brain from the recordings on the scalp. The problem of recovering the distribution of neural activity inside the brain given the potential distribution on the scalp is underdetermined, with many possible source activities that can account for the same scalp potential distribution, leading to a myriad of methods available [28], [29]. Nevertheless, the results quite often aid in characterizing neurological disorders [30], [31], [32], [33]. In this methodology, we opted for fitting dipoles to independent components [34]. This method estimates the locations of a small number of dipoles and their activities directly from the data, as opposed to perhaps more popular and easier to use source imaging methods such as minimum norm estimation [35] and its variations, or sLORETA [36], which both assume fixed locations and sometimes fail when multiple source activities are considered [37]. The fMRI is then used to provide the coordinates of the region of interests (ROIs) in the relevant default mode network (which differ from subject to subject), and only the activities of the dipoles that are closest to these sources are selected for further analysis.

The next step is to compute time-varying connectivity between all pairs of ROIs, in the five different EEG frequency bands, using five different functional connectivity measures. Functional connectivity has been chosen rather than effective connectivity, because, in this way, connectivity is computed from the data, and not from a model [14], [16]. We describe each connection by both the median of this time-varying functional connectivity and by several indicators of structural connectivity obtained from DTI. We then find in which circumstances the functional network correlates the most with the structural network by observing the largest positive or negative correlation coefficients.

The aim of this study is to present a detailed description of this novel multimodal analysis using fMRI, DTI and EEG. A pilot application in a small sample of children with ASD is provided. The significant correlations between EEG and DTI features suggest the feasibility of using such an approach to infer structural properties from EEG data. If the results of this pilot study can be confirmed in a larger sample of subjects, the proposed model could be applied for the longitudinal non-invasive analysis of autistic brain and for the monitoring of the effect of the treatment on neural connectivity.

## II. Subjects and Data Acquisition

Three children diagnosed with ASD have been recruited for this pilot study at IRCCS 'Stella Maris Foundation' (Pisa, Italy), a tertiary care referral centre and University Hospital. We used a multi-informant perspective on clinical data collection by using professional observation or interview (ADOS-G [38], Griffiths Scales [39] and Vineland Adaptive Behaviour Scales II [40]) and parent reports (Child Behaviour Checklist [41] 1.5-5 and 6-18 versions, Social Communication Questionnaire [42] and Parenting Stress Index [43]). Children's inclusion criteria were:
- Age between 4 and 8 years;
- Meeting criteria for ASD, according to the Diagnostic and Statistical Manual of Mental Disorders, 5th Edition [5];
- Autism classification confirmed by the ADOS-G (administered from psychologists trained in the use of ADOS in a research setting);
- Full Scale Intelligence Quotient (FSIQ) of 80 or higher on the Wechsler Scales.

The research protocol for Magnetic Resonance (MR) and EEG assessment has been defined and evaluated for approval by the institutional review board of the Stella Maris Clinical Research Institute for Child and Adolescent Neurology and Psychiatry, Pisa, Italy. Functional magnetic resonance imaging and diffusion tensor imaging scans were acquired on a 1.5T GE scanner at the same hospital. Children were instructed to lie quietly inside the scanner while remaining awake, with open eyes, during the experiment, and to perform no specific cognitive exercise. As the Human Connectome Project has shown that the spatial patterns of resting-state fMRI correlations are stable across multiple 'resting' states, such as eyes-open, eyes-closed, fixation and sleep, the "open-eyes" condition was chosen to increase the compliance of the patients inside the scanner [44].

Before the MR acquisition, children performed a fake session inside a scanner simulator ('Zero Tesla'), which has all the characteristics of a real MR scanner. It allows children to familiarize themselves with the environment and the acoustic noise of the scanner before the real session, in order to reduce head motion during image acquisition. The acquisition protocol includes the following sequences:
- T1-weighted fSPGR (Voxel size $1\times1\times1$ mm$^3$, reconstruction diameter 250 mm);
- Diffusion Tensor Imaging (TR=10 s, TE=92 ms Voxel size $3\times3\times3$ mm$^3$, reconstruction diameter 240 mm, 30 gradient directions, b-value 1000 s/mm$^2$);
- Resting-state fMRI (TR=2.4 s, TE=50 ms, flip angle=90°, Voxel size $3.75\times3.75\times3$ mm$^3$, slice thickness=4mm, reconstruction diameter 240 mm).

In a separate session, EEG data was acquired from the subjects. EEG signals were obtained in the same experimental condition as fMRI data, i.e. resting state, eyes-open condition. In this pilot study, the data has been recorded using an ENOBIO system with 19 dry electrodes employing the 10-20 electrode placement system and a sampling frequency of 500 Hz [18]. The lengths of the recordings were 694.5s, 602.7s and 917.2s, but the length of the third was reduced from 987.2s due to heavily corrupted with artefacts.

III. OBTAINING REGIONS OF INTEREST FROM FMRI

Functional brain connectivity is evaluated in a resting-state condition. By measuring intrinsic, or spontaneous, fluctuations in the blood oxygen level dependent (BOLD) signal in the absence of an externally imposed task it is possible to characterize anatomical connectivity and spontaneous communication across large-scale networks in the brain [45]. In particular, the default-mode network (DMN) has special relevance to autism because many of the same areas are activated by tasks requiring complex emotional and social processes, theory of mind, and self-referential thought [46].

Processing of the fMRI images was carried out using FMRI Expert Analysis Tool (FEAT), (part of FMRIB Software Library (FSL), FMRIB, Oxford, UK) [47]. The fMRI data are corrupted by motion and noise, therefore pre-processing steps are required to enhance the quality of the image. Motion correction was obtained using MCFLIRT, which implements an affine transformation among images within the same time-series. Spatial smoothing using a Gaussian kernel of full-width at half maximum (FWHM) of 5 mm and high-pass temporal filtering (Gaussian-weighted least-squares straight line fitting, with $\sigma = 50.0$s) were applied to increase the signal to noise ratio (SNR). The BET tool was used for non-brain tissue removal and grand-mean intensity normalization of the entire 4D dataset by applying a single multiplicative factor. In addition, fMRI data was registered to high-resolution structural space (T1-weighted fSPGR image) using affine registration. Registering functional to anatomical data is useful both for looking at single subject results, i.e. overlaying statistical results onto the subjects structural scan, and as a precursor for warping subjects into standard space for multiple-subject analysis.

The identification of the DMN was achieved using the Independent Component Analysis (ICA) method. In fMRI analysis functional connectivity is assessed in terms of the linear correlations of blood-oxygen-level-dependent (BOLD) time course data between two or more spatially remote regions. Seed-based or clustering techniques have previously been used for this purpose, but ICA is nowadays the most popular method for extracting multiple distributed functional connectivity patterns of neural activity from whole-brain fMRI time-series without a priori specification of a temporal profile or spatial layout for the BOLD responses or any specific region as 'seed' for the statistical characterization of the patterns. Spatial ICA has been shown to reliably extract a statistical image of the DMN network in single subject and groups, under several fMRI experimental settings.

The ICA was applied in the probabilistic version [48] as implemented in Multivariate Exploratory Linear Decomposition into Independent Components (MELODIC), part of FSL. Temporal concatenation ICA was performed across all functional datasets from each subject using automatic dimensionality estimation. The whitened observations were decomposed into sets of vectors, which describe signal variation across the temporal domain (time-courses) and across the spatial domain (maps) by optimizing for non-Gaussian spatial source distributions using a fixed-point iteration technique. The independent components (ICs) were obtained using a cluster level $z = 2.3$ and a threshold

$p<0.05$. The ICs were examined by visual inspection. Noisy components with extreme power spectra, sudden jumps in intensity, motion and susceptibility artefacts due to air signal were excluded. The other components were evaluated for their similarity with the DMN. The DMN definition is based on literature as the network comprising the Superior Frontal Gyrus/Medial Prefrontal Cortex (SFG, mPFC), the Posterior Cingulate Cortex (PCC), the Precuneus, and the Lateral Parietal Cortex [49]. For each subject one or two ICs have been identified as representing the DMN. In Figure 1 the IC related to the DMN identified for the third subject has been shown as a representative example.

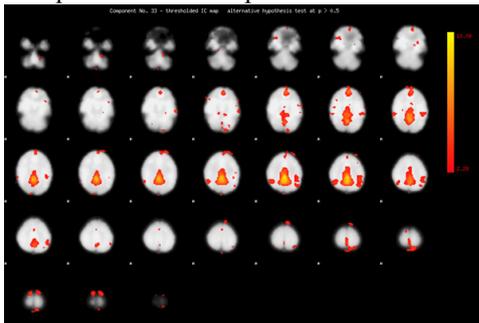

Figure 1: Independent Component identified as related to the default mode network.

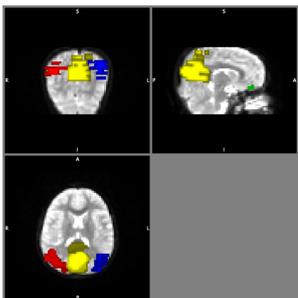

Figure 2: Region of interest masks creation on the brain regions of the default mode network. **Red**: Right Parietal Cortex; **Blue**: Left Parietal Cortex; **Yellow**: Precuneus/Posterior Cingulate Cortex; **Green**: Medial Prefrontal Cortex.

The activation related to the DMN was then superimposed on the T1-weighted anatomical image, which was previously co-registered with the functional image. For each of the brain regions composing the DMN, a ROI mask is created. The identification of the ROI was guided by the Atlas Tool inside FSL. The Atlas Tool provided a *probabilistic* atlas of macroscopic anatomy, which means that each structure in the atlas is represented as a standard space image with values from 0 to 100, according to the cross-population probability of a given voxel being in that structure. Once the correct structure was identified, a mask was realized for that structure for brain activity above a certain threshold ($p<0.05$). Figure 2 shows an example of ROI masks obtained for the Precuneus/Posterior Cingular Cortex, Right Parietal Cortex, Left Parietal Cortex and Superior Frontal Gyrus/Medial Prefrontal Cortex in the first subject. These brain regions are the fMRI 'sources' identified in the resting state analysis.

For each of these brain regions the $x$, $y$ and $z$ coordinates were obtained. These are the coordinates in the subject reference system. In order to compare the activation among different subjects it is more suitable to have coordinates in a common reference system. The standard MNI space was selected as template. First the pre-processed fMRI image was registered to the standard MNI space using the *flirt* tool of FLS that performs an affine registration. The calculated affine transformation that registers the input (fMRI image) to the reference (MNI template) is saved in a 4×4 affine matrix. This matrix was used to register the masks realized on the selected brain area to the MNI template. The new co-registered masks were visualized superimposed to the MNI template and for each mask the coordinates were extracted. In conclusion, for each brain region or 'source', the coordinates in the subject and in the standard space were obtained. Table 1 gives the ROIs in the MNI space for all three subjects under study. Figure 3 summarizes the process applied for sources identification and coordinates extraction.

TABLE 1: COORDINATES OF THE FMRI SOURCES BELONGING TO THE DEFAULT MODE NETWORK IN MNI SPACE FOR ALL THE SUBJECTS

|  | Subject 1 | | | Subject 2 | | | Subject 3 | | |
|---|---|---|---|---|---|---|---|---|---|
| *ROI name* | x | y | z | x | y | z | x | y | z |
| PCUN/PCC | -2 | -94 | 38 | -2 | -72 | 38 | -2 | -38 | 38 |
| SFG/mPFC | 12 | 26 | 8 | 4 | 34 | -10 | 2 | 42 | 12 |
| LPC | -26 | -104 | 26 | -24 | -82 | 46 | -46 | -64 | 32 |
| RPC | 24 | -104 | 26 | 24 | -82 | 46 | 44 | -62 | 32 |

*PCUN/PCC – Precuneus/Posterior Cingulate Cortex; SFG/mPFC – Superior Frontal Gyrus/Medial Prefrontal Cortex; LPC – Left Parietal Cortex; RPC – Right Parietal Cortex;*

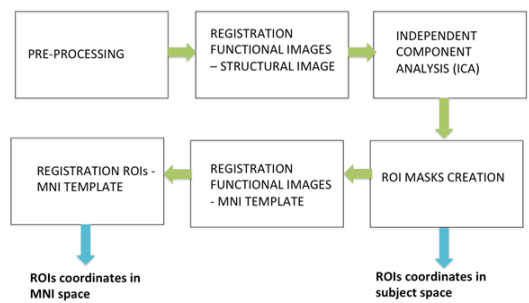

Figure 3: The process chain for deriving fMRI ROI or source coordinates for each subject.

## IV. OBTAINING STRUCTURAL CONNECTIVITY FROM DTI

Anatomical connectivity is obtained through DTI analysis, which is the only method that presently allows measurement of white matter fibre orientation in the human brain in-vivo [50]. The white matter fibre tracts are actually large bundles of axons that interconnect the grey matter processing areas both within and across hemispheres. In diffusion MRI, the quantity measured relates to the three-dimensional organization of the tissue. In this context, we apply the DTI analysis to compute anatomical connections among the different fMRI sources identified in section III.

DTI images were first pre-processed using the FSL [47] software package. First, data quality assessment was performed by looping through the individual images to check for gross artefacts, such as signal dropouts and interleave artefacts caused by sudden subject motion. No participants had to be excluded based on these data checks. Then, for each subject, all images including diffusion weighted and b0 were corrected for eddy current induced distortion and subject motion effect using FDT (FMRIBs Diffusion Toolbox) [51]. Brain mask was created from the first b0 image using BET (Brain Extraction Tool) [52]. For each subject a suitable fractional threshold was selected so that the mask includes all the brain structures. The fractional anisotropy (FA) is a value between 0 and 1, where the threshold must be set between these two limits. Usually, for children, a threshold between 0.45 and 0.5 is optimal for a good masking of the brain.

Once pre-processing was completed, the diffusion tensor vectors were calculated using ExploreDTI software (http://www.exploredti.com/). The images corrected for eddy current distortions and motions in FSL were imported in ExploreDTI for tractography. We chose this software because it allows the setting of several parameters in the reconstruction algorithm and it gives a good performance in terms of accuracy of the reconstructed tracts [21]. First images were converted from the NIfTI data format to the Matlab .mat format in order to make them suitable for the analysis.

Once converted, tracts were computed using a whole brain tractography algorithm. More specifically, the deterministic approach implemented by Basser *et al.* [53] was selected for computation of tracts. Deterministic refers to the propagation modality of the reconstructed tract that is determined by the direction of maximum diffusivity of the voxel that is considered. The first step of this algorithm consists in the association of the major eigenvector with the tangent to a curve (the putative fibre path). Then the curve is estimated by stepping repeatedly in the direction of the tangent. A cubic method was used to interpolate the tensors estimated at each voxel's level. Some other parameters were tuned on the data analysed in order to obtain an optimal reconstruction of tensors: uniform 2 mm seed point resolution, 0.5 mm step size, FA tracking threshold range between 0.2 and 1, fibre length range between 20 and 500, angle threshold of 30 degree and an FA tract termination threshold of 0.2. The performance of the algorithm was verified by checking the correct orientations and structure of tracts.

The calculation of tracts connecting the ROIs, which in our case are the sources selected with the fMRI analysis, was performed using TrackVis (http://www.trackvis.org/). In order to import tracts obtained with ExploreDTI in TrackVis an in-house Matlab program was used, which takes as inputs the diffusion tensor images and the tractography files, both in .mat format, and gives as output a tractography file in the format used by the TrackVis Software. This program was realized in such a way that also the information about length, FA and mean diffusivity (MD) of each single tract would be transferred to the TrackVis software. With this transformation the whole brain tractography, i.e. all the fibre tracts present in each subject's brain, is available for quantitative analysis.

In contrast to whole-brain tractography, locally-constrained tractography makes use of ROI-based Boolean operations, such as specifying volumes through which fibres must or may not pass. As a result, locally-constrained tractography offers higher sensitivity and greater control for tracking selected fibres of interest. In this study, what is needed is a locally-constrained tractography in which the ROIs are the sources identified by the fMRI analysis.

In order to visualize the ROI masks realized in the fMRI analysis on DTI images it is necessary to register them to the diffusion space. First the pre-processed fMRI image was registered to the b0 DTI image using the *flirt* tool of FLS that performs an affine registration. The calculated affine transformation that registers the input (fMRI image) to the reference (DTI image) is saved in a 4×4 affine matrix. This matrix was used to register the masks to the diffusion space. Once co-registered, the ROIs were imported in TrackVis for the reconstruction of the tracts between the ROIs. DTI fibre tractography was used to estimate the likely connections between the 4 ROIs. Tracts that did not end in or pass through both ROIs were discarded. In Figure 4 an example of the tracts connecting the ROIs for the third subject is shown. It is evident that for this subject there are connections between the Precuneus/Posterior Cingulate Cortex and all other ROIs, but no connection between the Right and Left Parietal Cortex has been found.

TABLE 2: SUMMARY OF THE INFORMATION ABOUT THE TRACTS CONNECTING THE SOURCES FOR ALL SUBJECTS

| Connection | Subject 1 | | | Subject 2 | | | Subject 3 | | |
|---|---|---|---|---|---|---|---|---|---|
| | NoT | LEN | VOL | NoT | LEN | VOL | NoT | LEN | VOL |
| PCUN/PCC - LPC | 137 | 263 | 19.41 | 14 | 243.6 | 5.508 | 15 | 112.8 | 3.38 |
| PCUN/PCC - RPC | 154 | 260.7 | 22.04 | 23 | 237.7 | 5.346 | 10 | 100.8 | 2.54 |
| LPC - RPC | 172 | 270.1 | 20.82 | 17 | 267.1 | 13.35 | - | - | - |
| PCUN/PCC – SFG/mPFC | - | - | - | - | - | - | 9 | 239 | 5.91 |

*PCUN/PCC – Precuneus/Posterior Cingulate Cortex; LPC – Left Parietal Cortex; RPC – Right Parietal Cortex; SFG/mPFC – Superior Frontal Gyrus/Medial Prefrontal Cortex; NoT – Number of Tracts; LEN – Average Tract Length (mm); VOL – Connection Volume (mm³)*

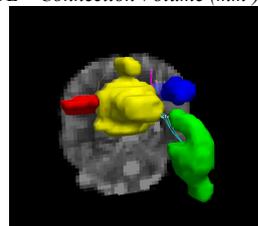

Figure 4: Tracts connecting the sources for one subject. These are: *orange, top left* - between the Precuneus/Posterior Cingulate Cortex (yellow) and the Right Parietal Cortex (red); *pink, top right* - between the Precuneus/ Posterior Cingulate Cortex and the Left Parietal Cortex (blue); *cyan, bottom right* - between the Precuneus/ Posterior Cingulate Cortex and the Superior Frontal Gyrus (green).

For the existing tracts some statistics are extracted in particular: the volume, the length and the number of fibres. These measures thus provide an indication of the strength of the connections between sources. Figure 5 summarizes the entire process of DTI analysis, tracts identification and statistics extraction. Table 2 summarizes the information about structural connectivity obtained from this study.

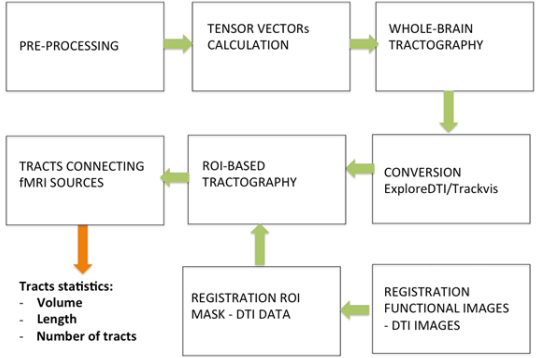

Figure 5: The process chain for deriving statistics about tracts connecting fMRI sources from DTI analysis.

## V. SOURCE LOCALIZATION FROM EEG, GUIDED BY fMRI REGION OF INTERESTS

### A. Independent Component Analysis for Source Localization

Each EEG electrode records a superposition of activities originating from multiple regions in the brain. Estimating these activities and their locations is known as the *source-localization* problem [28], [29]. A large number of assumptions need to be made to solve this. Typically, a source is modelled as an equivalent current dipole, described by 6 parameters: three *x*, *y*, *z* parameters to localize it in space, two $\theta$ and $\varphi$ parameters to describe its orientation and one *d* parameter to describe its strength. For the purpose of the present work, we assume that only the dipole strength varies in time, while the other parameters remain fixed. Therefore $d = d(t)$ will be called dipole activity in the remainder of the paper. Given a model describing the geometry of the head and how currents propagate through the different types of tissue, if the parameters of a dipole are known, the activity it produces on the scalp can then be computed, also known as the *forward problem*. For a given number of *P* dipoles and *N* electrodes, this can be summarized as

$$\overline{x}(t) = G\overline{d}(t) + n. \quad (1)$$

where $\overline{x}(t)$ is the vector of *N* time-varying measurements on the scalp, $\overline{d}(t)$ is the vector of *P* dipole activities, *G* is an *N*×*P* matrix relating the two, called the *lead-field matrix*, and *n* accounts for the noise in the measurements. The lead-field matrix is computed from the head model and its elements depend on the positions and orientations of the dipoles. Finding the dipole activities, their position and orientations given the measurements and the head model is called the *inverse problem*, and it involves searching for the $\hat{G}$ and $\hat{\overline{d}}$ that would minimize the cost function

$$J = \left\| \overline{x} - \hat{G}\hat{\overline{d}} \right\|. \quad (2)$$

This problem is undetermined, with many possible solutions, and solving it requires making further assumptions about the distribution of brain activity. To find a possible solution, we will use the temporal version of ICA [54] and thus make the assumption that the dipole activities are statistically independent.

If $\overline{x}(t)$ is the vector of *N* EEG signals, ICA will yield an *N*×*N* un-mixing matrix *W* and the corresponding *N* independent components $\overline{s}(t)$, such that

$$\overline{s}(t) = W\overline{x}(t) \quad (3)$$

The activities of the independent components $\overline{s}(t)$ can thus be viewed as an approximation of the dipole activities (if *N* dipoles are considered), so $\overline{s}(t)$ is an approximation of $\overline{d}(t)$, but note that no physical significance is actually associated with these activities at this point. Connectivity computation would be possible, but it is not known what each activity corresponds to. To associate locations to the dipoles, the unmixing matrix *W* needs to be analyzed. Its inverse is called the *mixing matrix*. In the mixing matrix, column *i* contains the weights with which the signal from component *i* contributes to each of the scalp electrodes. If its elements are represented on a scalp plot termed component topography one can, for example, recognize components corresponding to blinking artefacts by their increased weights in the frontal region. Apart from artefact rejection, source localization can also be performed as *dipole fitting* [34] that aims to fit each component topography with an equivalent current dipole, characterized by position and moment. In this context, the $W^{-1}$ can be considered as an approximation of *G*. The elements of *G* are in fact nonlinear functions of the dipole locations and moments, computed based on the head model and the locations of the electrodes. If *G* is known, then finding, for each column, the location and moment corresponding to a dipole, can be solved as a nonlinear optimization problem. In other words, for each component, dipole fitting finds the parameters of the dipole that produces the scalp map closest to the component topography, by minimizing residual variance in an optimization problem.

Functions from EEGLAB [55] were used to carry out the processing. Prior to ICA computation, the data was first band-pass filtered between 0.5 and 45 Hz using EEGLAB's FIR filter, which uses a heuristic to set the order and returned order 3301 and transition band width 0.25 Hz. The ICA algorithm used was *Runica* [55]. *Runica* is the EEGLAB implementation of the *infomax* algorithm, which achieves maximum statistical independence by maximizing the joint entropy of the output of a neural network, by employing a gradient descent algorithm. The algorithm was called with learning rate set to heuristic rather than the default of 0.001, increasing stability at the expense of convergence speed. The 'extended' option was also set, which allows the detection of components with a negative kurtosis.

The *Dipfit* plugin for EEGLAB was then used to fit dipoles to the scalp topographies of the independent components. This requires setting up a head model first. The head model used is the three-shell boundary element model of the standard MNI brain. Since the EEG cap uses the standard 10-20 system, the electrode coordinates in the MNI space are readily available and no co-registration between the head model and electrode locations is necessary. Fitting dipoles occurs in two steps. The *first step* is a coarse fitting, to find the best initial conditions for the subsequent nonlinear optimization. For this, fixed dipoles are assumed at each point on a 3-D grid of dipole positions from which 272 locations inside the head are retained. The scalp topography produced by each individual dipole acting alone is then computed using the head model. Next, each component topography is compared to all 272 different projections and the best match is selected, as measured by the residual variance of fitting the dipole projection to the component topography. The *second step* is a nonlinear iterative fitting of 6 dipole parameters (position and moment) to each component, where the cost function to minimize is the residual variance, and the initial conditions are given by the results of the previous step.

Ignoring the dipole moments, each independent component activity now effectively has an estimated source location attached. For example, for the first subject, the 19 dipole locations attached to the 19 independent components are shown in Figure 6.

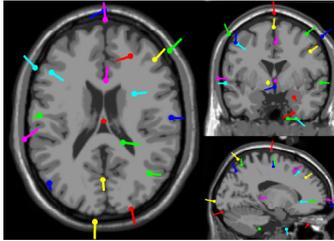

Figure 6. Example of dipoles that were fit to the 19 independent components obtained from EEG, plotted on a template MRI image. Clockwise: top view, coronal view and sagittal view.

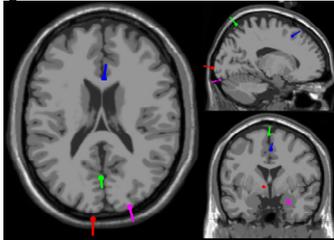

Figure 7. Example of dipoles that were selected among the 19 as being the ones that are the nearest in Euclidean space to the fMRI ROIs.

### B. Selecting Independent Components using fMRI

Dipole fitting returns the coordinates in normalized MNI space of at most as many dipoles as the number of EEG channels, associated with the activities of independent components (ICs). However, not all these components correspond to neural activity, and not all of them are of interest. If information about the ROIs is available from fMRI, then it is possible to select which components should be further considered for connectivity analysis by considering the locations of their associated dipoles. We therefore propose the selection of the independent components used in connectivity analysis based on the proximity of their dipoles to the ROIs, as measured by Euclidean distance. In other words, each fMRI location $i$ will be matched with the $n^{th}$ dipole, where

$$n = \arg\min_j \left( dist\left( ROI_i, dipole_j \right) \right) \quad (4)$$

Having selected the dipoles of interest, their 4 associated independent components are retained for connectivity computation, whereas the rest are discarded. This is done for each subject using the MNI coordinates of the 4 ROIs provided (Precuneus/Posterior Cingulate Cortex, Left Parietal Cortex, Right Parietal Cortex, Superior Frontal Gyrus/Medial Prefrontal cortex). For example, the best matching dipoles are plotted for the first subject in Figure 7, while the complete list of coordinates is shown in Table 3.

TABLE 3: COORDINATES OF THE EEG SOURCES THAT ARE THE BEST MATCHES FOR THE fMRI ROIs

| ROI name | Subject 1 | | | Subject 2 | | | Subject 3 | | |
|---|---|---|---|---|---|---|---|---|---|
| | x | y | z | x | y | z | x | y | z |
| PCUN/PCC | -0.65 | -69.48 | 68.45 | 11.7 | -69.84 | 68.99 | 0.9 | -70.29 | 67.92 |
| SFG/mPFC | 1.98 | 19.93 | 48.34 | -52.53 | 49.92 | -3.3 | 23.63 | 72.28 | -0.6 |
| LPC | -7.94 | -107 | -2.81 | -31.52 | -94.37 | 21.55 | -63.89 | -49.17 | 35.43 |
| RPC | 24.73 | -95.64 | -21.86 | 39.56 | -85.36 | 32.43 | 34.88 | -88.46 | 35.19 |

*PCUN/PCC – Precuneus/Posterior Cingulate Cortex; SFG/mPFC – Superior Frontal Gyrus/Medial Prefrontal Cortex; LPC – Left Parietal Cortex; RPC – Right Parietal Cortex*

### VI. HIGH TEMPORAL RESOLUTION IN THE FUNCTIONAL CONNECTIVITY FROM EEG

Time-varying connectivity can be obtained by sliding a window across the estimated source activities and computing the connectivity at each position of the window. To obtain the finest detail in recording the changes in connectivity, the window position has been incremented by only one sample.

If we define a window $w(t)$ of length $l$, such that

$$w(t) = \begin{cases} 1, & t \in [0,l) \\ 0, & t \notin [0,l) \end{cases}. \quad (5)$$

The windowed source activity $i$ is obtained by sliding this window will be $s_i(t)w(t-\tau)$, for all $\tau = \{0,\cdots,T-l\}$, where $T$ is the length of the signal, such that only full length windows are considered. The signals in each window are normalized to zero mean and unit variance.

Functional connectivity is then computed between all pairs of source activities, to obtain a time-varying connectivity matrix $A(\tau)$, for which an element $a_{ij}(\tau)$ represents the connectivity strength between sources $i$ and $j$ at window position $\tau$.

$$a_{ij}(\tau) = F\left( s_i(t)w(t-\tau), s_j(t)w(t-\tau) \right) \quad (6)$$

where, $F = F(x(t), y(t))$ is a functional connectivity measure that takes two signals and returns a real value for the degree of their synchronization. Here, $a_{ij}(\tau) = a_{ji}(\tau)$, since functional connectivity has been used, which does not capture the directionality of the signal interactions. Two classes of functional connectivity measures are employed here to estimate the connectivity between all pairs of source activities, from those reviewed in [56].

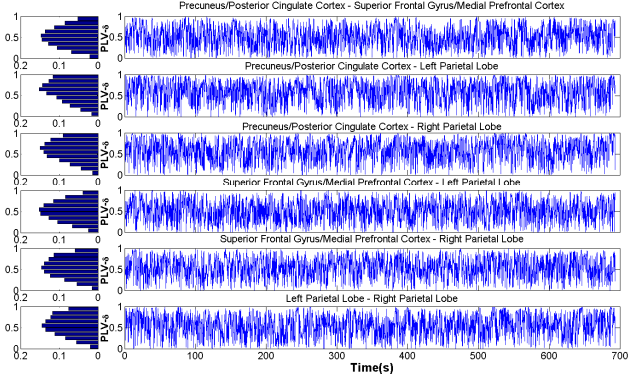

Figure 8. Example of time-varying connectivity between all pairs of regions for one of the subjects, together with their normalized histograms. Connectivity has been computed with PLV in the δ band.

*Phase synchronization measures* aim to quantify the temporal stability of the phase difference $\Delta\varphi_{xy}(t)$ between the two signals *x(t)* and *y(t)*. *Phase-locking Value* (*PLV*) averages the representations of the phase difference on the unit circle across all available time points and then takes the magnitude of the result.

$$PLV(x(t), y(t)) = \left|\left\langle \exp(j\Delta\varphi_{xy}(t)) \right\rangle\right| \quad (7)$$

where, $\langle . \rangle$ indicates averaging or expectation over time. *Phase-lag Index* (*PLI*) averages the sign of the phase difference across all time points, effectively discarding distributions centered on zero, which is considered to be indicative of volume conduction.

$$PLI(x(t), y(t)) = \left|\left\langle sign(\Delta\varphi_{xy}(t)) \right\rangle\right| \quad (8)$$

The *RHO index* (also known as the phase entropy) is a measure of how much the entropy of the distribution of the phase values deviates from the entropy of the uniform distribution.

$$RHO(x(t), y(t)) = (S_{max} - S)/S_{max} \quad (9)$$

where, *S* is the Shannon entropy of the distribution of $\Delta\varphi_{xy}(t)$ and $S_{max}$ is the entropy of a uniform phase distribution).

*Coherency-based measures* have evolved from the notion of coherency, which is the cross-spectral density of the two signals ($P_{xy}$) normalized by the product of their auto power spectral densities ($P_{xx}, P_{yy}$).

$$K_{xy}(f) = \frac{P_{xy}(f)}{\sqrt{P_{xx}(f)P_{yy}(f)}} \quad (10)$$

*Coherence* (*COH*) takes the square magnitude of this quantity, and the *imaginary part of coherency* (*iCOH*) is also used as a precaution against volume conduction. Since COH and iCOH are functions of frequency, we average these over the frequency bands of interest.

$$COH(x(t), y(t)) = \frac{1}{F_2 - F_1 + 1} \sum_{f=F_1}^{F_2} |K_{xy}(f)|^2$$

$$iCOH(x(t), y(t)) = \frac{1}{F_2 - F_1 + 1} \sum_{f=F_1}^{F_2} \text{Im}(K_{xy}(f)) \quad (11)$$

Computing connectivity between all pairs of sources will give 6 unique time-varying connectivity strengths from the 4 sources. These are estimated using Matlab functions adapted from the Hermes Toolbox [56], implementing the functional connectivity measures – *PLV, PLI, RHO, COH, iCOH*. They are applied to a sliding window of the 4 estimated sources activities, with length 500 samples (1 second), and overlap between successive window positions of 499 samples, yielding connectivity time courses that are 1 second (one window length) shorter than the original signals.

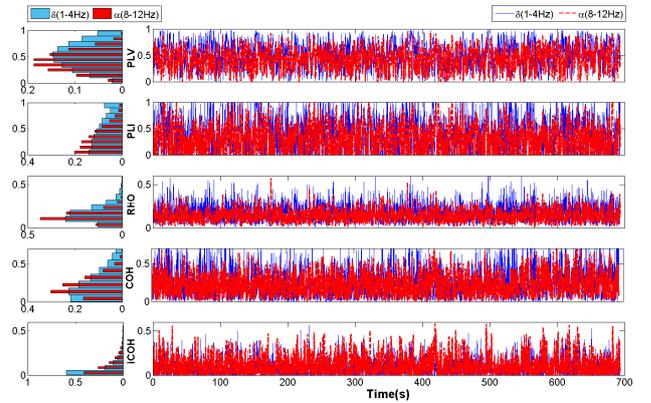

Figure 9. Example of time-varying connectivity between Precuneus/Posterior Cingulate Cortex and Superior Frontal Gyrus/Medial Prefrontal Cortex shown for one of the subjects, together with their normalized histograms. For comparison, results obtained with all five connectivity measures and in the frequency bands δ and α are shown.

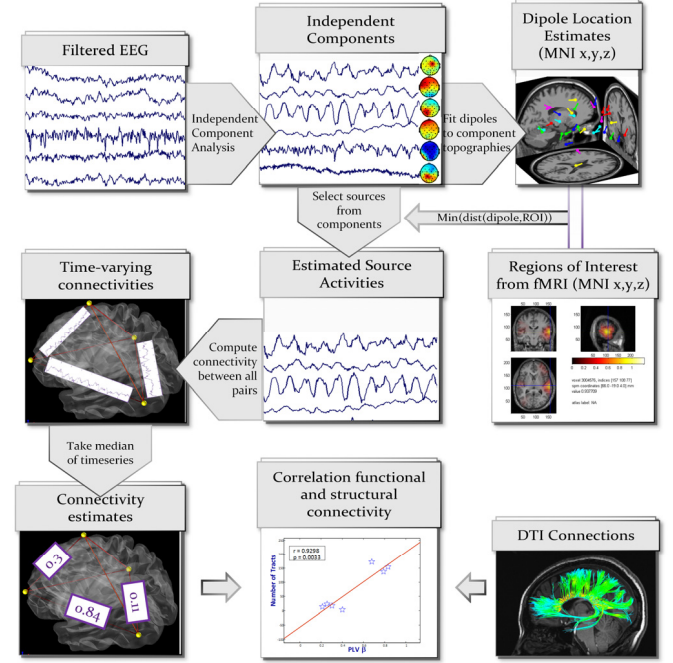

Figure 10. Overview of the steps taken to obtain functional connectivity from EEG and fMRI and to correlate that information with structural connectivity obtained from DTI.

TABLE 4: SUMMARY OF THE STRUCTURAL AND FUNCTIONAL CONNECTIVITY FOR ALL THE CONNECTIONS THAT CONTAIN PHYSICAL WHITE MATTER TRACTS

| Subject | S1 | S1 | S1 | S2 | S2 | S2 | S3 | S3 | S3 |
|---|---|---|---|---|---|---|---|---|---|
| ROI 1 | PCUN/PCC | PCUN/PCC | LPC | PCUN/PCC | PCUN/PCC | LPC | PCUN/PCC | PCUN/PCC | PCUN/PCC |
| ROI 2 | LPC | RPC | RPC | RPC | RPC | RPC | LPC | RPC | SFG/mPFC |
| NoT | 137 | 154 | 172 | 14 | 23 | 17 | 15 | 10 | 9 |
| LEN | 263.01 | 260.74 | 270.08 | 243.64 | 237.69 | 267.05 | 112.8 | 100.75 | 239 |
| VOL | 19.41 | 22.04 | 20.82 | 5.508 | 5.346 | 13.346 | 3.38 | 2.54 | 5.91 |
| PLV-δ | 0.46 | 0.44 | 0.47 | 0.63 | 0.60 | 0.57 | 0.55 | 0.65 | 0.60 |
| PLV-θ | 0.34 | 0.33 | 0.44 | 0.61 | 0.65 | 0.51 | 0.40 | 0.52 | 0.37 |
| PLV-α | 0.33 | 0.33 | 0.34 | 0.74 | 0.75 | 0.65 | 0.42 | 0.55 | 0.37 |
| PLV-β | 0.21 | 0.26 | 0.30 | 0.79 | 0.83 | 0.68 | 0.27 | 0.64 | 0.40 |
| PLV-γ | 0.25 | 0.34 | 0.41 | 0.84 | 0.87 | 0.73 | 0.36 | 0.64 | 0.51 |
| PLI-δ | 0.36 | 0.34 | 0.36 | 0.42 | 0.41 | 0.44 | 0.42 | 0.49 | 0.49 |
| PLI-θ | 0.25 | 0.24 | 0.40 | 0.28 | 0.39 | 0.32 | 0.25 | 0.26 | 0.26 |
| PLI-α | 0.25 | 0.24 | 0.27 | 0.30 | 0.39 | 0.41 | 0.26 | 0.24 | 0.24 |
| PLI-β | 0.17 | 0.20 | 0.23 | 0.18 | 0.28 | 0.28 | 0.24 | 0.15 | 0.14 |
| PLI-γ | 0.21 | 0.22 | 0.42 | 0.20 | 0.15 | 0.21 | 0.32 | 0.14 | 0.14 |
| RHO-δ | 0.15 | 0.14 | 0.15 | 0.20 | 0.15 | 0.15 | 0.15 | 0.17 | 0.16 |
| RHO-θ | 0.11 | 0.11 | 0.14 | 0.21 | 0.17 | 0.14 | 0.12 | 0.17 | 0.12 |
| RHO-α | 0.12 | 0.12 | 0.11 | 0.29 | 0.21 | 0.18 | 0.12 | 0.18 | 0.12 |
| RHO-β | 0.04 | 0.05 | 0.06 | 0.30 | 0.30 | 0.19 | 0.05 | 0.18 | 0.08 |
| RHO-γ | 0.05 | 0.07 | 0.09 | 0.36 | 0.35 | 0.22 | 0.07 | 0.19 | 0.12 |
| COH-δ | 0.16 | 0.15 | 0.21 | 0.34 | 0.34 | 0.24 | 0.24 | 0.31 | 0.24 |
| COH-θ | 0.14 | 0.13 | 0.20 | 0.44 | 0.46 | 0.33 | 0.23 | 0.33 | 0.19 |
| COH-α | 0.12 | 0.12 | 0.18 | 0.62 | 0.66 | 0.50 | 0.19 | 0.40 | 0.17 |
| COH-β | 0.14 | 0.17 | 0.23 | 0.77 | 0.82 | 0.62 | 0.21 | 0.54 | 0.31 |
| COH-γ | 0.18 | 0.26 | 0.32 | 0.83 | 0.87 | 0.69 | 0.27 | 0.55 | 0.41 |
| iCOH-δ | 0.04 | 0.03 | 0.07 | 0.04 | 0.05 | 0.04 | 0.04 | 0.05 | 0.05 |
| iCOH-θ | 0.05 | 0.05 | 0.11 | 0.07 | 0.11 | 0.10 | 0.06 | 0.06 | 0.06 |
| iCOH-α | 0.05 | 0.06 | 0.09 | 0.07 | 0.15 | 0.15 | 0.06 | 0.06 | 0.05 |
| iCOH-β | 0.08 | 0.09 | 0.13 | 0.08 | 0.12 | 0.12 | 0.11 | 0.07 | 0.06 |
| iCOH-γ | 0.10 | 0.10 | 0.20 | 0.10 | 0.08 | 0.10 | 0.15 | 0.08 | 0.08 |

*S1 – Subject 1; S2 – Subject 2; S3 – Subject 3; PCUN/PCC – Precuneus/Posterior Cingulate Cortex; LPC – Left Parietal Cortex; RPC – Right Parietal Cortex; SFG/mPFC – Superior Frontal Gyrus/Medial Prefrontal Cortex; NoT – Number of Tracts; LEN – Average Tract Length (mm) ; VOL – Connection Volume (mm³). ROI 1 and ROI 2 are the names of the two ROIs involved in the connection.*

Computations are performed in the five canonical frequency bands, $\delta$ (1-4 Hz), $\theta$ (4-8 Hz), $\alpha$ (8-12 Hz), $\beta$ (12-30 Hz) and $\gamma$ (30-45 Hz), since previous studies relating EEG and resting-state networks obtained from fMRI found relevant information in all frequency bands [57], [58]. To summarize, this gives 6 time-varying connectivity strengths, obtained through 5 connectivity measures, in 5 different bands. This is done for all the three ASD subjects. An example of such a full estimate of time-varying connectivity for the second subject and all possible connections is shown in Figure 8, where connectivity was computed using *PLV* in the $\delta$ band. Figure 9 shows a comparison of estimating the time-varying strength for a single connection with all 5 different measures in frequency bands $\delta$ and $\alpha$, for the same subject.

Unlike functional connectivity, structural connectivity changes very slowly, and can be considered fixed in the present scenario. Therefore, in order to find a mapping between the structural connectivity and functional connectivity, the dynamic nature of the latter needs to be summarized to a single feature per connection. This is achieved by considering the distribution of the values in time and computing the median value.

The full analysis workflow is summarized in Figure 10, as follows:

*Step one*: Decompose the filtered multichannel EEG into independent components;
*Step two*: Fit dipoles to all scalp component topographies;
*Step three*: Obtain coordinates of ROIs from fMRI;
*Step four*: Select only those components whose dipoles are closest to the ROIs, based on Euclidean distance criteria;
*Step five*: Estimate time-varying connectivity between all pairs of ROIs by sliding a window through the sources activities and applying the synchronization measures described in equations (7)-(11), for the $\delta$, $\theta$, $\alpha$, $\beta$ and $\gamma$ bands;
*Step six*: Obtain structural connectivity information from DTI and find the features from functional connectivity that yield the best positive/negative correlation with the structural connectivity features.

## VII. CORRELATION ANALYSIS BETWEEN STRUCTURAL AND FUNCTIONAL CONNECTIVITY IN THE ASD SUBJECTS

All connections from all subjects are first aggregated together such that each connection, which can be described by 3 features of structural connectivity and 25 features of functional connectivity (5 connectivity measures × 5 frequency bands), is considered as a data point. A summary of the functional and structural connectivity information available is given in Table 4. Columns correspond to different connections, while rows correspond to different methods of describing these connections. The first three rows correspond to structural connectivity (number of tracts, length and volume from DTI), while the rest correspond to functional connectivity estimated from EEG in different ways in different frequency bands.

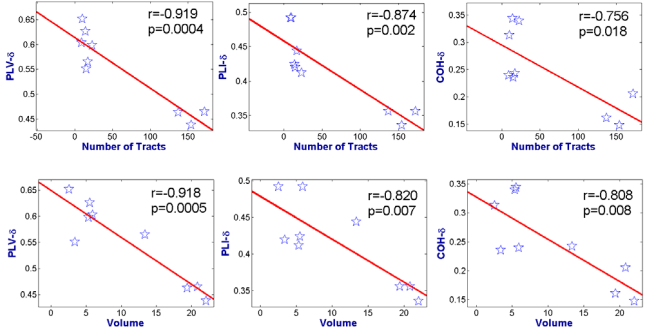

Figure 11. Plots of the best 6 correlations between features of structural and functional connectivity. The correlations Number of Tracts *vs.* PLV-$\delta$ and Tract Volume *vs.* PLV-$\delta$, shown in the first column, are statistically significant ($p<0.05$, Bonferroni corrected for $n=75$). The next four best correlations are also shown.

We then compute the Pearson correlation coefficient ($r$) corresponding to all 75 (= 3 structural features × 25 functional connectivity features) possible pairings of features from each type of connectivity. The $p$-values corresponding to the probability of the null hypothesis of zero correlation are also computed. We find one significant negative correlation between functional connectivity estimated from PLV in the $\delta$ band and the number of tracts, and one significant negative correlation between PLV in the $\delta$ band and tract volume ($p<0.05$, Bonferroni corrected for $n=75$ comparisons). These are the only combinations that pass this stringent statistical hypothesis test. It would, nevertheless, be informative to investigate when other high (positive/negative) values of the correlation coefficient are obtained. We find that the overall strongest 6 correlations, corresponding to a correlation coefficient $r<-0.75$ and an uncorrected $p$-value $p<0.05$, occur in the $\delta$ band, namely some of those in which functional connectivity was computed using PLV, PLI or COH. These are all negative correlations and are shown in Figure 11.

Whenever a functional connectivity measure is strongly linked to the number of tracts, it will also be linked to the tract volume. This occurs because the number of tracts and the tract volume are, in fact, also correlated among themselves ($r = 0.92$, $p = 0.0003$). Tract length did not correlate well with any of the functional connectivity measures: length *vs.* PLI-$\delta$ and length *vs.* PLV-$\delta$ were the best combinations, only reaching $r = -0.51$ and $r = -0.49$.

The significant correlations obtained suggest that it would be possible to estimate the underlying structural connectivity from the non-invasively measured EEG functional connectivity. All significant correlations are found to be negative. For example, the negative correlation found between PLV in the $\delta$ band and the number of tracts or volume would essentially mean that the stronger the physical connection between two activated regions of the brain, the lesser the phase coupling in the low frequency $\delta$ band oscillations. This might seem counter-intuitive, but the effect of an increase in the number of tracts on the phase characteristic of the filtering performed by the medium in which the signal propagates is unknown – and might very well decrease phase synchronization. This is just one of many possible explanations for the observation.

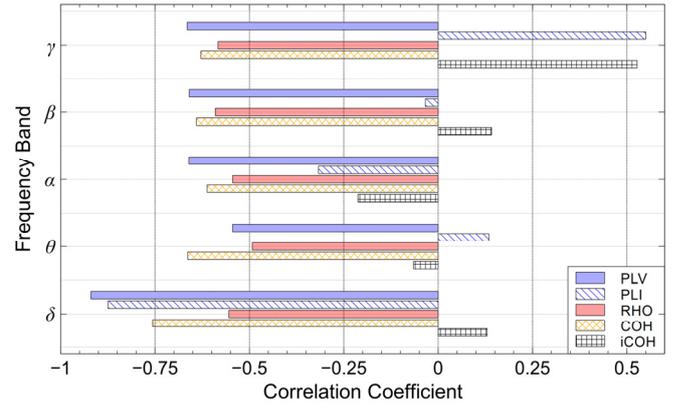

Figure 12. Correlation coefficients between the number of tracts and functional connectivity obtained using different measures, grouped by frequency band. The highest +/- correlation values occur when estimating functional connectivity in the $\delta$ band.

Figure 12, Figure 13 and Figure 14 show the values of the correlation between functional and structural connectivity grouped according to frequency bands, separately for each of the three measures of structural connectivity. It becomes apparent that the $\delta$ band shows the strongest links between function and structure, whereas no significant patterns emerge

in the higher frequencies. Resorting again to modelling the structural connections as filters, it would be straightforward to assume that the connections in this network simply low-pass filter the signals, but it isn't necessarily so. Since the measures of functional connectivity, we employ, rely on phase synchronization, a series of band-pass filters with a nonlinear phase characteristics that disrupts the phase synchronization everywhere but in the low frequencies would be sufficient to cause the same effect. The physical explanation of the band specific variation in the phase synchronization of source level brain activities is an open problem in neuroscience and could be explored as a complex mathematical model of large network of filters in a future research.

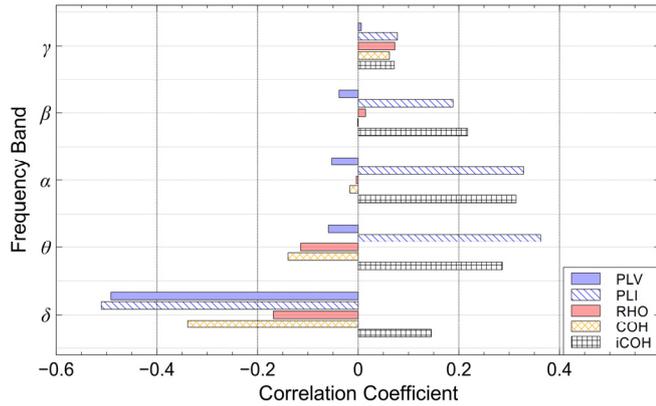

Figure 13. Correlation coefficients between the average tract length and functional connectivity obtained using different measures, grouped by frequency band. The highest +/- correlation values occur when estimating functional connectivity in the $\delta$ band.

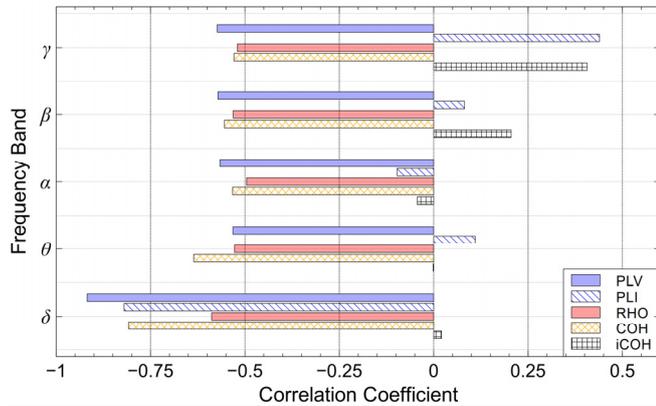

Figure 14. Correlation coefficients between the tract volume and functional connectivity obtained using different measures, grouped by frequency band. The highest +/- correlation values occur when estimating functional connectivity in the $\delta$ band.

VIII. DISCUSSION AND CONCLUSION

The methodology described in this paper requires multi-modal recording of EEG, fMRI and DTI. Previous studies have implemented multimodal imaging approaches to study brain structure and function (for a review please see [59]). These are mainly bimodal studies combining structural MRI with either EEG or fMRI, as a functional modality. Only few attempts have been performed to implement three-modal models e.g. [60][61][62] and rarely they have included DTI [63][64], [65]. To the best of our knowledge, none of these three-way approaches have been applied before for the investigation of autistic brain. Moreover, while the aim of these studies was to correlate different measure of brain connectivity to enhance insights on the neural correlates of a particular disease, the main aim of our study was to estimate the underlying connectivity based on EEG. For this purpose we used a wearable non-invasive EEG system that can be easily applied in children with ASD and potentially at home. Electroencephalography source imaging (ESI) is a powerful technique to localize the sources of the signals recorded on the scalp and to investigate the temporal dynamics of neural circuits. Combining EEG with other modalities provides priors that allow estimating the communication between sources at every moment in time [66]. As in previous ESI studies e.g. [67][68] we have used priors derived from resting-state fMRI, however in our studies we have also correlated the results the features extracted from DTI analysis. This is a novel ESI approach in which fMRI informed ESI was used to infer structural connectivity. Such an approach can give important insight into the autistic brain using a non-invasive methodology as the wearable EEG.

The main contribution of this work can be summarized as follows:

- Using the notion of dynamic functional connectivity, we have estimated the time varying interdependence between the activities of relevant sources in the autistic brain. Among many possible sources revealed by EEG source localization, the relevant ones have been selected as those that are closest to the fMRI-based regions of interest, in this case the ones in the DMN.

- Treating the time-varying connectivity between two sources as a random variable, we have analyzed its distribution and we have computed an expected value. This method of estimating functional connectivity has the advantage that it does not discard its intrinsic dynamic nature – and makes full use of the temporal resolution of the EEG.

- We found that the best way to capture the effect of the underlying structural connectivity of the DMN is by using PLV as a measure of estimating functional connectivity in the $\delta$ band. This shows the potential of estimating structural connectivity only from EEG functional connectivity. For example, one could use the slopes of the correlation plots in Figure 11 to reconstruct the number of tracts or tract volume, only using the EEG recording. Many different measures of signal interdependence have been reported in literature to estimate functional connectivity – from the subset we here report the ones that produce estimates closest to the underlying structural network.

- We also found that for the DMN, functional connectivity from fMRI and EEG correlates with structural connectivity most often in the $\delta$ band and the correlation is not that significant in other higher frequency bands.

One of the most intriguing findings of this study was the significant negative correlation between EEG connectivity in delta band and measures of structural connectivity, specifically tract length and tract volumes. Previous studies have found a significance between the functional connectivity in the DMN and $\delta$ power [69], [57]; which is consistent with our findings of significant correlations within this frequency band. Deficit in the DMN have been linked to mirror neurons dysfunctions, which in turn have been found associated to social impairments in ASD [70]. The $\delta$ band has been linked to learning, motivation and reward processes [71][72], as well as to memory encoding and retrieval [73]. In ASD both significant increase or decrease of $\delta$ power during resting state has been previously reported [74][75][76][10], suggesting an impairments of these processes in the disorder. More interestingly, $\delta$ abnormalities have been linked to a disconnection between gray and white matter [77]. These findings could agree with our finding of a negative correlation between $\delta$ connectivity and tract length and volumes. This correlation may indicate a possible compensatory mechanism in ASD. However it should be highlighted that the direction of impaired connectivity in ASD (i.e. hypo-connectivity *vs.* hyper-connectivity) is still controversial, with very heterogeneous findings like pointed questions has been recently termed as the 'autism connectivity chaos' [78].

Although the findings of our study seems promising for the study of connectivity in ASD, due to the preliminary nature of the study there are some limitations of the proposed methodology as well, which are left for the scope of future research e.g.

- Small sample size, which prevent a generalization of the results. Further studies would be needed to apply the proposed methodology on a larger sample,
- Lack of a typically developing sample group. Since in usual clinical practice typically developing subjects, especially children, do not undergo fMRI or DTI, validating the methodology on a typically developing group has been left as an open problem and might be considered as the scope of future research,
- Considering different set of ROIs for a wider spectrum of autistic subjects,
- Validating the methodology using high density EEG *vs.* the presently used low density pervasive/mobile EEG system,
- Assessing age, gender and treatment effects on a larger population of ASD children
- Testing other integration approaches of multimodal data.

Recently some modeling approaches have been introduced, which combine multimodal data, including functional and structural brain imaging data, to model spatio- and spectro-temporal brain data (STBD) [79]. In particular NeuroCube in [79], [80] is a spiking neural network combining structural, functional and, optionally, genetic data, which has been developed to model and understand STBD. The main advantage of this modeling approach is that it is suitable not only to estimate but also to predict brain activity. Some issues of the computational model approaches include the difficulty of modeling the diversity and emerging complexity of the nervous system, especially in diseased case [81]. In the future it would be interesting to compare our statistical approach for data integration with such a modeling approach.

Withstanding these limitations, from a practical perspective, the technique proposed in this paper could be applied to measure neural change in response to therapy in ASD individuals through an easy-to-use approach. Specifically, longitudinal studies that include pre- and post-treatment acquisitions may provide new insights on the neural mechanisms targeted in rehabilitative intervention and in addition, an objective measure of response to treatment.